\documentclass[fleqn,10pt]{wlscirep}

\usepackage{amsmath}
\usepackage{graphicx}
\usepackage{dcolumn}
\usepackage{bm}
\usepackage{float}
\usepackage{todonotes}

\title{Relativistic magnetic reconnection driven by a laser interacting with a micro-scale plasma slab}

\author[1,2,*]{Longqing Yi}
\author[3]{Baifei Shen}
\author[4]{Alexander Pukhov}
\author[1]{T\"unde F\"ul\"op}
\affil[1]{Department of Physics, Chalmers University of Technology, 41296 Gothenburg, Sweden}
\affil[2]{State Key Laboratory of High Field Laser Physics, Shanghai Institute of Optics and Fine Mechanics, Chinese Academy of Sciences, P.O. Box 800-211, Shanghai 201800, China}
\affil[3]{Department of Physics, Shanghai Normal University, Shanghai, 200234, China}
\affil[4]{Institut f\"ur Theoretische Physik I, Heinrich-Heine-Universit\"at D\"usseldorf, D\"usseldorf, 40225 Germany}

\affil[*]{corresponding author: longqing@chalmers.se}

\begin{abstract}

  \textbf{Magnetic reconnection is a fundamental plasma process
    associated with conversion of the embedded magnetic field energy
    into kinetic and thermal plasma energy, via bulk acceleration and
    Ohmic dissipation. In many high-energy astrophysical events,
    magnetic reconnection is invoked to explain the non-thermal
    signatures. However, the processes by which field energy is
    transferred to the plasma to power the observed emission are still
    not properly understood. Here, via 3D particle-in-cell simulations
    of a readily available (TW-mJ-class) laser interacting with a
    micro-scale plasma slab, we show that when the electron beams
    excited on both sides of the slab approach the end of the plasma
    structure, ultrafast relativistic magnetic reconnection occurs in
    a magnetically-dominated (low-$\beta$) plasma. The resulting
    efficient particle acceleration leads to the emission of
    relativistic electron jets with cut-off energy $\sim$ 12 MeV.  The
    proposed scenario can significantly improve understanding of
    fundamental questions such as reconnection rate, field dissipation
    and particle acceleration in relativistic magnetic reconnection.}
\end{abstract}

\begin{document}

\flushbottom
\maketitle
\thispagestyle{empty} Many high-energy astrophysical environments are
strongly magnetized, i.e.~the magnetic energy per particle exceeds the
rest mass energy, so that the magnetization parameter $\sigma\equiv
B_{0}^{2}/4\pi nm_{e}c^{2}\geq1$, where $B_{0}$ is the magnetic field
strength, $m_{e}$ is the electron mass, $n$ is the plasma density, and
$c$ is the vacuum light velocity. In such environments, magnetic
reconnection (MR), operating in the relativistic regime, plays a key
role in the transfer of large amount of magnetic to kinetic energy via
field dissipation \cite{s1,s15,s13}. Motivated by numerous
astrophysical observations, such as high energy emission in pulsars
\cite{s2}, cosmological gamma-ray bursts \cite{s7} and active galactic
nuclei jets \cite{s8,s9}, the study of relativistic MR has made rapid
progress in the last few decades through analytical
studies \cite{s18,s19} as well as 2D \cite{s68,s21,s23,s26} and 3D
\cite{s12,s27,s29,s30,s31} particle-in-cell (PIC) simulations. However, due to difficulties in
achieving the extreme magnetic energy densities that are required to
observe relativistic MR in laboratory environments, previous
experimental studies investigated mainly the non-relativistic regime
($\sigma<1$). These include experimental observation of MR in tokamaks
\cite{s32} or dedicated experiments, such as MRX \cite{s33}.

High-intensity laser-plasma interaction
\cite{s34,s37,s35,s36,s38,s39,s40} is a promising way to break through
the relativistic limit as the energy densities that
can be achieved by high-intensity laser facilities
worldwide\cite{s41,s42} are rising rapidly. These facilities
provide an important platform for the study of relativistic MR and
therefore attract extensive attention. In the typical laser-driven MR
experiments, two neighboring plasma bubbles are created by
laser-matter interaction. The opposite azimuthal magnetic fields arise
due to the Biermann battery effect ($\nabla n\times\nabla T_{e}$,
where $T_{e}$ is the electron temperature) \cite{s44} and reconnect in
the midplane as they are driven together by the frozen-in-flow of bulk
plasma expansion. Although most of the previous studies are
focused on the non-relativistic limit, it has been reported recently
that relativistic MR conditions can be achieved with such a scenario
in laboratory environments \cite{s46}.

In the laser-driven reconnection studies based on the Biermann battery
effect \cite{s34,s37,s35,s36,s38,s39,s40,s46,s65,s66}, due to the
oppositely directed magnetic fields that need to be compressed
together by the plasma thermal flows, the ratio of the plasma thermal
and magnetic energy, $\beta$, is high ($\beta>1$). As a result, the plasma is not  magnetically dominated and it is therefore not clear what role
MR plays in terms of energy balance. Recently it has been reported
that magnetically dominated MR can be achieved by a double-turn
Helmholtz capacitor-coil target \cite{s69}, but this approach is very
difficult to extend to the relativistic regime because it requires a
kJ-class laser system. Thus, in spite of the remarkable progress that
has been made, relativistic MR in low-$\beta$ environments
($\beta<1$), which is closely related to the interpretation of many
space plasma measurements and astronomical observations \cite{s51},
has not been thoroughly studied.

In this paper, we propose a novel experimental setup based on the
interaction of a readily available moderately intense (TW-mJ-class)
laser with a micro-sized plasma slab, resulting in ultrafast
relativistic MR in the low-$\beta$-regime. By irradiating a
micro-sized plasma slab with a high-intensity laser pulse, it is
possible to deposit the laser energy in a very small volume, hence
significantly reducing the laser energy (by at least 1-2 orders of
magnitude comparing with the results reported in Ref.\cite{s46})
required to achieve relativistic MR regime ($\sigma>1$). Meanwhile,
since the magnetic field lines are not driven by the plasma thermal
flow, the magnetically-dominated regime can be accessed. Using 3D PIC
simulations, a comprehensive numerical experiment is presented, which
demonstrate the MR event observed in the proposed scheme have a
significant effect on the whole system. It leads to fierce
(0.1-TW-class) field dissipation and highly-efficient particle
acceleration, which covers 20$\%$ of the total energy transition. The
enormous magnetic tension gives rise to intense relativistic jets that
have so far been absent in other relativistic MR studies based on
laser-plasma interaction \cite{s46,s52,s53}.

Due to these features, the proposed scenario is promising to provide
an important platform to study the non-thermal signatures and energy
transition in relativistic MR. In addition, we present other MR
signatures including quantified agyrotropy peaks in the diffusion area
\cite{s55}, and out-of-plane quadrupole field structures \cite{s56}.
With recent advances in laser pulse cleaning techniques
\cite{s57,s58,s59} and micro-target manufacturing \cite{s60}, the
proposed scenario can be straightforwardly implemented in experiments.

\begin{figure}[!b]
\centering
\includegraphics[width=15.5cm]{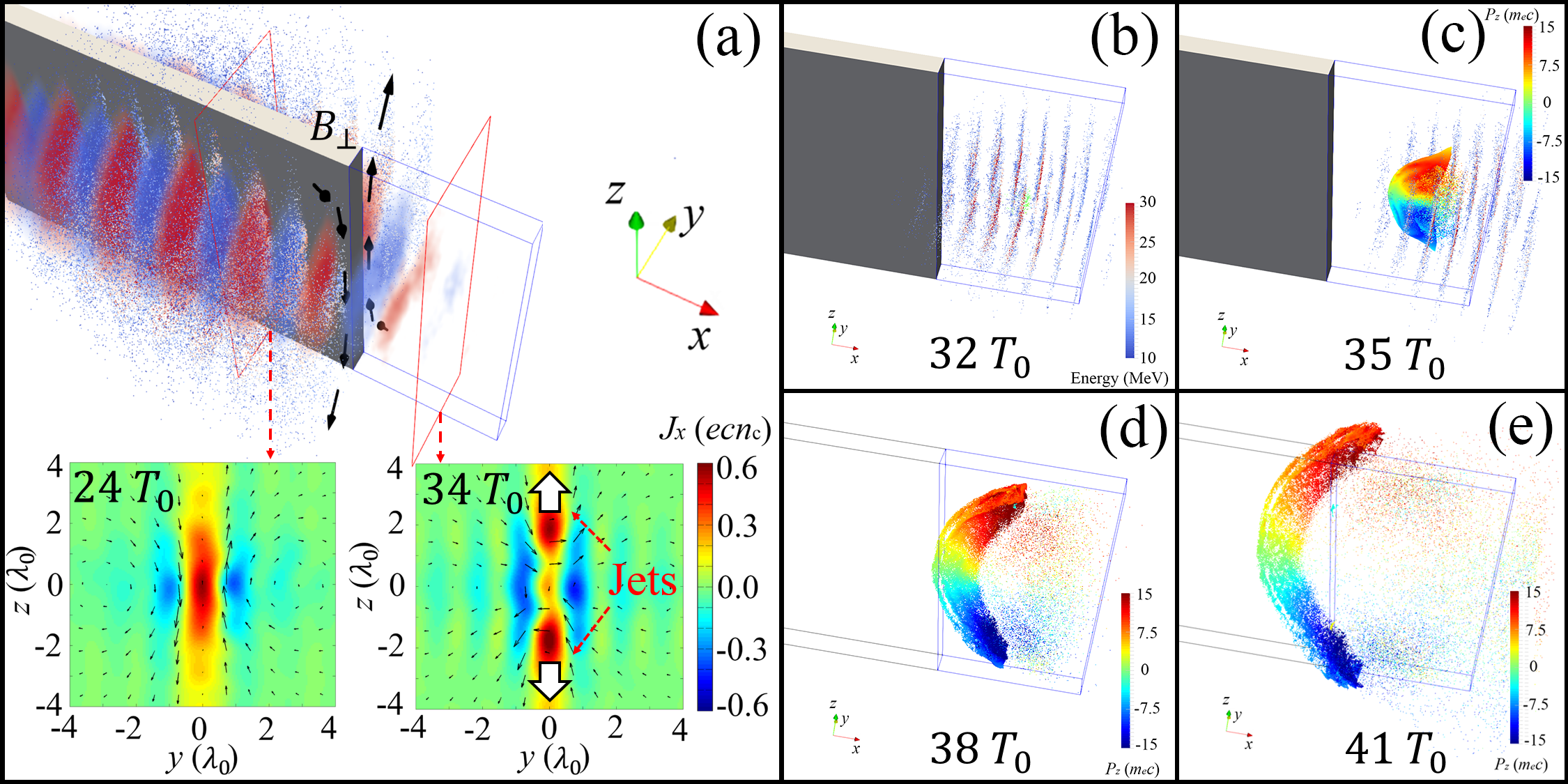}
\caption{\textbf{Schematic of the proposed setup and relativistic jets
    generation.} (a) A moderately high-intensity laser pulse
  ($a_{0}=5$) propagates along the $x$-direction, and is splitted in
  half by a micro-sized plasma slab. The laser drives two energetic
  electron beams on both sides of the plasma surfaces, which generates
  100 MG level opposing azimuthal magnetic fields in the
  middle. Ultrafast magnetic reconnection is observed as the electron
  beams approach the coronal region (the area within the blue box,
  where the plasma density decreases exponentially) at the end of the
  slab.  The two insets below show the transverse magnetic field (black
  arrow) and longitudinal electric current density (color) at the
  cross-section marked by the red rectangle (separated by 10
  $\lambda_{0}$) at simulation times t = 24 $T_0$ and t = 34 $T_0$,
  respectively. (b-e) Generation and evolution of the relativistic jet
  resulting from MR at times 32, 35, 38 and 41 $T_0$,
  respectively. The rainbow color-bar shows the transverse momentum
  $P_z$ of the jets formed by the background plasma electrons and the
  blue-red color-bar in (b) shows the energy of the electron bunch that is
  driven by the laser pulse.}
\label{fig:1}
\end{figure}

\subsection*{Generation of relativistic jets.}

A sketch of the simulation setup is shown in Fig.~1(a). A linearly
polarized (in $y$ direction) laser with normalized laser intensity
$a_{0}=eE_{0}/m_{e}c\omega_{0} = 5$ (intensity $\sim$ 10$^{19}$
W/cm$^{2}$) propagates along  the $x$-axis, where $E_{0}$ is the laser amplitude,
$\omega_{0} = 2\pi c/\lambda_{0}$ and $\lambda_{0} = 1\mu$m are the
frequency and wavelength of the laser, respectively. The laser spot
size is 4 $\lambda_{0}$ and the duration is 15 $T_{0}$, where
$T_{0} \approx 3.3$ fs is the laser cycle. A plasma slab with
thickness (in the laser-polarizing direction) $d=1\lambda_{0}$
and length (in the laser propagation direction) $L=20\lambda_{0}$
splits the laser pulse in half. The main part of the slab has
an uniform density of $20n_{c}$, where $n_{c}=m_{e}\omega_{0}^{2}/4\pi
e^{2}$ is the critical density. At the end of the structure
(coronal region, represented by the area within the blue-box framework
in Fig.~1), the density drops exponentially as $x$ increases (scale
length $l=2\lambda_{0}$).  As the laser pulse sweeps along the
slab, it drives two energetic electron beams on both sides of the
plasma surface \cite{s61}. These electron beams are typically
overdense (i.e.~$n_{b}>n_{c}$) \cite{s62} and capable of generating
100 MG level opposing azimuthal magnetic fields in the middle, as
shown by the black arrows in Fig.~1(a). Detailed parameters of
the simulation can be found in \textbf{Methods}.

In the early stage (i.e.~before the electron beams reach the corona),
MR does not occur because a much stronger return current is
excited inside the slab, which separates the antiparallel magnetic
fields on each side of the slab, and the magnetic energy remains constant (after
the initial rise due to the laser-matter interaction) during this
period. As the electron beam approaches the end of the structure, the
plasma density decreases rapidly so that the electron number in the
local plasma becomes insufficient to form a return current that is
strong enough to separate the magnetic fields. Therefore,
due to Amp\`{e}re's force law, the electron beams on both sides
attract each other and flow into the mid-plane coronal plasma. The
magnetic field lines that move with the electron beams are pushed
together and reconnect. The reconnection magnetic field in the corona
is approximately 100 MG. An X-point magnetic field topology is
observed as shown by the right-bottom inset of Fig.1(a).

As the field topology changes, the explosive release of magnetic
energy results in the emission of relativistic jets as shown by
Fig.~1(b-e). These jets are formed by the background plasma electrons
in the corona, which start to appear at approximately $t = 32T_{0}$
[Fig.~1(b)] and acquire relativistic energies within a few laser
cycles. They propagate backwards ($-x$ direction) towards the exhaust
region of the reconnection site ($\pm z$ direction). The backward
longitudinal momentum stems from efficient acceleration due to
magnetic energy dissipation as will be discussed in the remainder of
this work. The electrons in the jets distinguish themselves from the
rest of background electrons that are heated by the laser pulse in
this region by (i) remarkably higher energies (top 0.2$\%$ on the
electron spectra, with mean energy $\bar{E}\sim$ 4.7 MeV), (ii)
considerably small y-divergence (i.e. $\theta_{y}=|P_{y}|/P\sim0.1$,
where $P$ is the electron momentum and $P_{y}$ denotes its
$y$-component), (iii) forming a dense electron beam, where the density
is $n_{jet} \sim 5 n_{c}$ initially, but decreases rapidly due to
dispersion in the $z$-direction. The total charge of the jets is
approximately 0.2 nC.

\subsection*{Pressure tensor agyrotropy and observation of field-line rearrangement.}

\begin{figure}[!b]
\centering
\includegraphics[width=15.5cm]{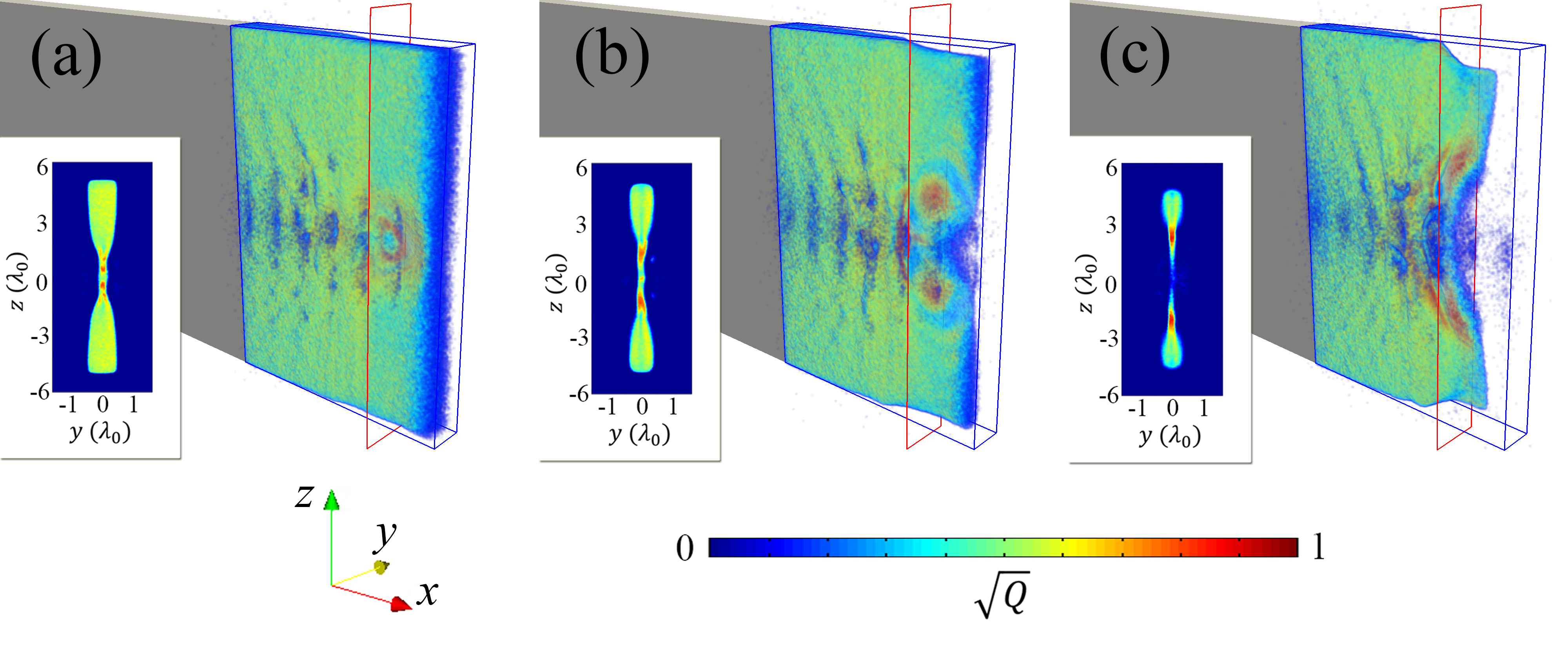}
\caption{\textbf{Gyrotropy quantification
at different times.} $\sqrt{Q}$ in the coronal plasma at
  simulation time $t = 32 T_{0}$(a), $33 T_{0}$(b), and $34
  T_{0}$(c). The insets show the value of $\sqrt{Q}$ at the
  cross-section with longitudinal coordinate $x = 26\lambda_{0}$,
  which is marked by the red rectangles in (a-c).}
\label{fig:2}
\end{figure}

In order to locate the reconnection site in the corona, we calculate
the scalar measure of the electron pressure tensor gyrotropy that was
suggested recently by Swisdak \cite{s55},
\begin{align}
\vspace{-10pt}
\begin{split}
Q=\frac{\mathcal{P}_{12}^{2}+\mathcal{P}_{13}^{2}+\mathcal{P}_{23}^{2}}{\mathcal{P}_{\perp}^{2}+2\mathcal{P}_{\perp}\mathcal{P}_{\parallel}}.
\end{split}
\vspace{-10pt}
\end{align}
Here
$\begin{pmatrix}
\mathcal{P}_{\parallel} & \mathcal{P}_{12} & \mathcal{P}_{13} \\
\mathcal{P}_{12} & \mathcal{P}_{\perp} & \mathcal{P}_{23} \\
\mathcal{P}_{13} & \mathcal{P}_{23} & \mathcal{P}_{\perp}
\end{pmatrix}$
is the electron pressure tensor
\begin{align}
\vspace{-10pt}
\begin{split}
\mathbb{P}=m\int(\mathbf{v}-\mathbf{\bar{v}})(\mathbf{v}-\mathbf{\bar{v}})fd^{3}\mathbf{v}
\end{split}
\vspace{-10pt}
\end{align}
transformed into a frame in which the diagonal components are in
gyrotropic form, i.e.~one of the coordinate axes points in the
direction of the local magnetic field and the others are oriented such
that the final two components of the diagonal of $\mathbb{P}$ are
equal (see the Appendix in Ref.~\cite{s55} for details).
$f(x,y,z,\mathbf{v})$ is the distribution function at position
$(x,y,z)$ and velocity $(\mathbf{v})$, and $\mathbf{\bar{v}}$ is the
mean velocity. It should be noted that although Eq.~(2) is the
non-relativistic definition of the electron pressure tensor, the
relativistic effects are not expected to have a significant influence
here since more than 99$\%$ of the electrons in the corona are
non-relativistic, namely $\gamma-1<1$ as can be seen in Fig.~4(c),
where $\gamma = 1/\sqrt{1-(v/c)^{2}}$ is the Lorentz factor of the
electrons. In general, for gyrotropic electron pressure tensors $Q=0$,
and the maximum departure from gyrotropy is $Q=1$. High values of $Q$
usually identify regions of interesting magnetic topology such as
separatrices and X-points in magnetic reconnection.

Figure~2 shows that the space and time where $Q$ reaches its
peak in the 3D PIC simulation coincides with the appearance of the
electron jets shown in Fig.~1(b-c). The slab is
significantly pinched during the interaction \cite{s63}, which results
in a higher inflow velocity and thus a faster reconnection rate.
Knowing the location of the reconnection site allows one to calculate the
magnetization parameter. By substituting the electron density and magnetic
field for the region with peak agyrotropy, one obtains the maximum
magnetization parameter $\sigma_{m}\approx25$. Therefore we conclude
that the observed MR is in the relativistic regime.

It is generally understood that the jets are accelerated out of the
reconnection region by the magnetic tension forces
($\textbf{\emph{T}}^{m}=(\textbf{\emph{B}} \cdot
\nabla)\textbf{\emph{B}}/4\pi$) of the newly-connected,
strongly-bent magnetic field lines. In Fig.~3, we plot the magnetic
fields at $x=26\lambda_{0}$, and the $z$-component of
the magnetic tension $\emph{T}^{m}_{z}$ at the midplane
($y=0$). A quadrupole longitudinal magnetic field pattern emerges as
the MR occurs, which is indicative of Hall-like reconnection
\cite{s56}, where the reconnection rate is significantly enhanced due
to decoupling of electron and ion motion. The amplitude of $B_{x}$ is
20 MG, which is $\sim$20$\%$ of the reconnected magnetic field.

\begin{figure}[!t]
\centering
\includegraphics[width=13.5cm]{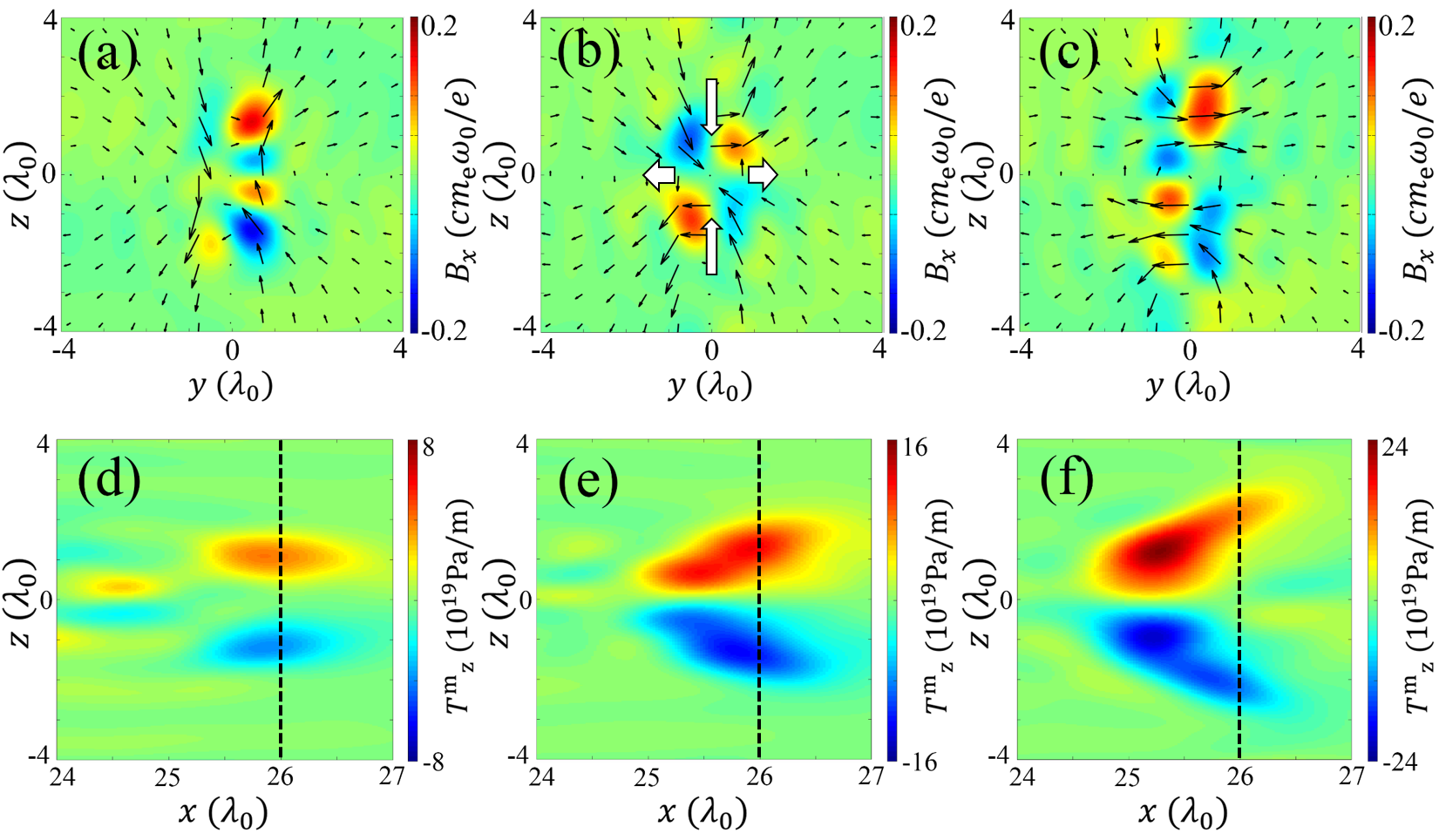}
\caption{\textbf{Evolution of magnetic fields and magnetic
  tension force during the reconnection.} (a-c) Static magnetic fields
  (frequency below 0.8$\omega_0$) and (d-f) $z$-component of magnetic tension force
  at simulation time t = 32 T$_0$ (a) \& (d), 33 T$_0$ (b) \& (e), and 34 T$_0$ (c) \&
  (f). In (a-c) the transverse ($B_y$, $B_z$) and longitudinal ($B_x$)
  components of magnetic field are presented by the black arrows and
  color, respectively. The bold white arrows in (b) show the inflow
  (horizontal) and outflow (vertical) electric currents that result
  from Hall reconnection. The black-dashed lines in (d-f) mark
  the cross-section where the corresponding magnetic fields (a-c) are
  shown.}
\label{fig:3}
\end{figure}

Figure~3 illustrates that an enormous magnetic tension force is
generated in the corona because of the reconnection. The pressure
exceeds the relativistic light pressure ($P = 2I/c$ with
$a_{0}\sim1$) within one laser wavelength, which exerts a strong
compression of the electron jets that results in the observed
high-density emission. Moreover, the shape of magnetic tension in
Fig.~3 shows the dynamics of the newly-connected field lines
strengthening and relaxing, during which
the energy is transferred to the plasma particles. The phenomenon
is consistent with the shape and emission direction of the
jets that we observed in Fig.~1(c-e).

\subsection*{Magnetic energy dissipation and particle acceleration.}

In order to gain a deeper understanding of energy transfer in the
relativistic MR, we now focus on the field dissipation process. The
observed dissipation power is of the order of 0.1-TW, which results in
a highly efficient energy transfer from the magnetic fields to the
kinetic energy of plasma.  Figure~4(a) is a graphic demonstration of
3D field dissipation, where the work done by the longitudinal electric
field per unit volume and unit time (i.e.~$E_{x}J_{x}$) is
presented. The blue chains on both sides of the slab show the process
of the laser-driven electron-beams losing energy to the static
electric field, which is sometimes referred to as ``enhanced target
normal sheath acceleration'' \cite{s64}. Meanwhile in the midplane,
one can see that the energy flows are directed in the opposite direction
as MR occurs, i.e.~the electrons in the coronal plasma \emph{collectively}
extract energy from the (reconnection associated) electric fields and thus
gain kinetic energy. To further clarify the dissipation process, we
also conducted a comparison simulation which is presented in the
\textbf{Supplemental Materials}.

\begin{figure}[!b]
\centering
\includegraphics[width=13.5cm]{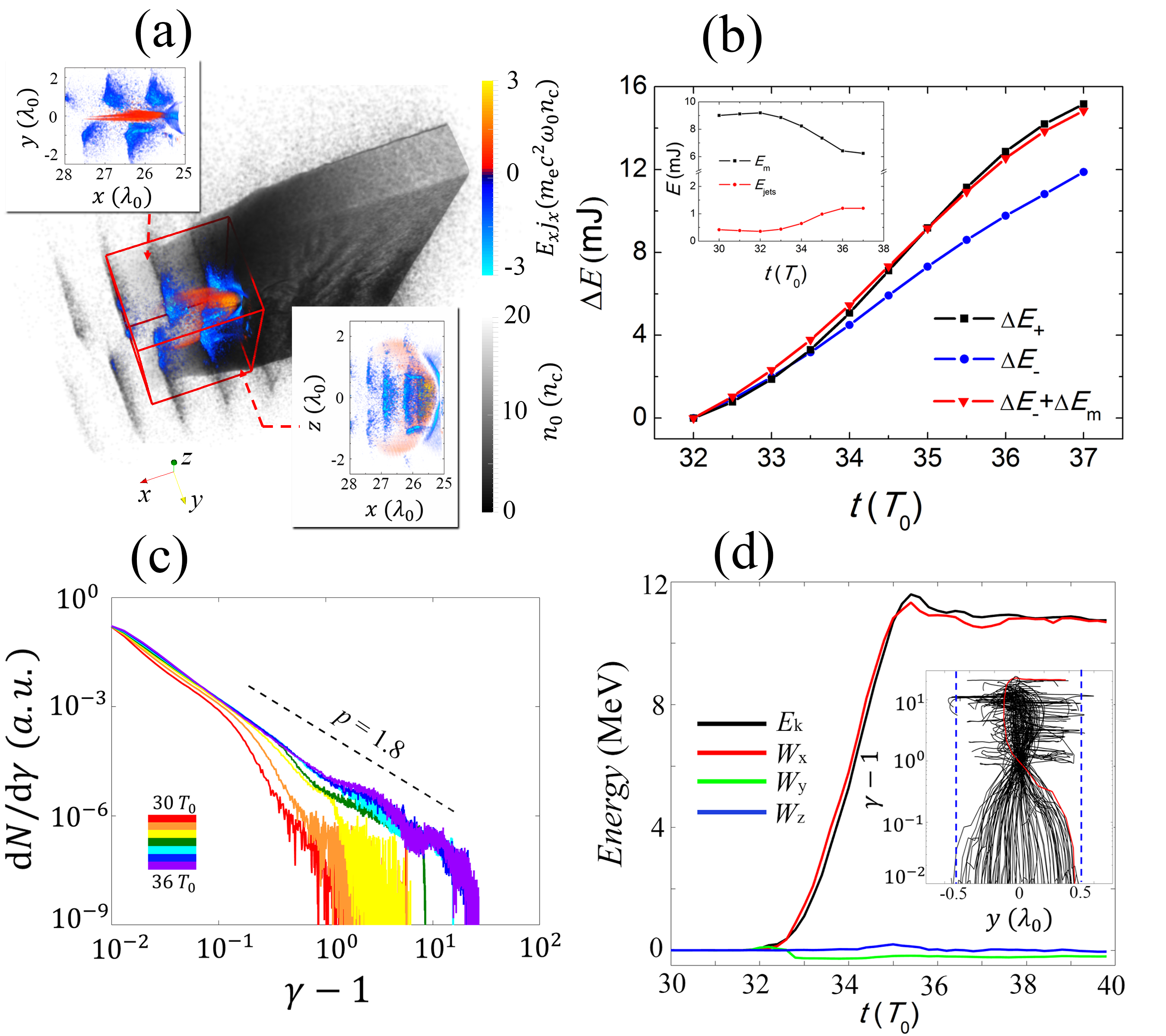}
\caption{\textbf{Magnetic energy dissipation and the energization
  of non-thermal electrons.} (a) Field dissipation ($E_{x}j_{x}$) and
  electron density at $t = 33T_{0}$ in the corona, the insets represent
  the top and side views of $E_{x}j_{x}$ in the reconnection site (marked by
  the red box). (b) Time dependence of total energy increase in electrostatic fields,
  electrons in the corona, and protons ($\Delta E_{+}$), energy reduction of
  electromagnetic fields and other electrons ($\Delta E_{-}$) as well as the total
  energy reduction that includes magnetic field dissipation ($\Delta E_{-}$+$\Delta E_{m}$),
  inset shows the evolution of static magnetic energy $E_{m}$ and total kinetic energy
  of electron jets. (c) Coronal electron spectra from 30 $T_{0}$ to 36 $T_{0}$.
  (d) The temporal evolution of the kinetic energy ($E_{k}$) and the work done by each
  electric field component ($W_{x}$, $W_{y}$, and $W_{z}$) for one representative
  electron. The inset plane shows the phase-space trajectory ($\gamma-1$ plotted vs $y$)
  of the total 100 tracked electrons, where the blue dashed line marks the boundary
  of plasma slab and the trajectory in red  represents the case shown in (d).
  }
\label{fig:4}
\end{figure}

To further understand the role that  relativistic MR plays in the
energy transfer process, we calculated the energy change for each
component in the simulation, i.e.~laser pulse (frequency $\geq
0.8\omega_{0}$), static electric/magnetic field (frequency $<
0.8\omega_{0}$), and kinetic energy of electrons and protons), and the
results are shown in Fig.~4(b). Let $\Delta E_{+}$ denote the total
energy increase in the electrostatic field, coronal electrons and protons,
$\Delta E_{-}$ denote the energy reduction in the laser pulse and
other electrons (mostly laser-driven electron beams), and $\Delta
E_{m}$ represent the energy loss in the static magnetic fields. As one
can see, a significant contribution to the total energy transfer comes
from the annihilation of magnetic fields due to relativistic MR, which
accounts for $\sim20\%$ of the total energy gain. Also, Fig.~4(b)
indicates the static magnetic field loses $\sim$ 3.0 mJ energy in 5
laser cycles, i.e.~an efficient magnetic energy dissipation with power
$\sim$ 0.18 TW is obtained in the simulation. In addition, the
evolution of the static magnetic energy and the kinetic energy of the electron
jets are plotted in the inset of Fig.~4(b). A good
time-synchronization is observed between the magnetic field
dissipation and electron acceleration.

In Fig.~4(c), we plot the electron energy spectrum in the corona from
$30T_{0}$ to $36T_{0}$. One can see that as the reconnection occurs, the
released magnetic energy is transferred to the non-thermal electrons. The
total electron kinetic energy in the corona has increased by a factor
of 4 during the reconnection and a hard power-law electron energy
distribution $dN/d\gamma\propto1/\gamma^{p}$ is obtained with
index $p\approx1.8$. In addition, according to the electron energy
spectrum at 30 $T_{0}$, the low-energy part (with $\gamma-1<0.1$,
contains 98.8$\%$ of the total charge) has a temperature around 6 keV,
given the reconnection magnetic field $\sim$ 100 MG and electron
density $\sim 1n_{c}$, the plasma $\beta$ is approximately only 0.02.

To reveal the mechanism of particle acceleration, we track 100
electrons that attain the highest energy during the MR, and display
one representative case in Fig.~4(d), where the electron kinetic
energy and work done by each component of the electric fields are
plotted as a function of time. It illustrates that the electric field
associated with reconnection at the X-point is primarily responsible for
the energization of the electrons. This is supported by the trajectory
of the tracked electrons in phase space [shown by the inset of Fig.~4(d),
where the trajectory of the representative electron for the energy
plotting is drawn in red], which shows that almost all the electrons
gain their energies within a narrow plane ($-0.1\lambda_{0}<y<0.1\lambda_{0}$)
adjacent to the X-point. In this region, since the magnetic field vanishes
due to the reconnection, the electrons become unmagnetized and can be
accelerated freely. Moreover, as indicated by the horizontal lines in
the phase space, once the electrons escape from this narrow plane, the
acceleration process stops immediately. Recent studies \cite{s65,s66}
have pointed out that the randomness of the electron injection and
escape from the acceleration region give rise  to the observed
power-law energy distribution as shown in Fig.~4(c).

\subsection*{Outlook.}

\begin{figure}[!b]
\centering
\includegraphics[width=8.5cm]{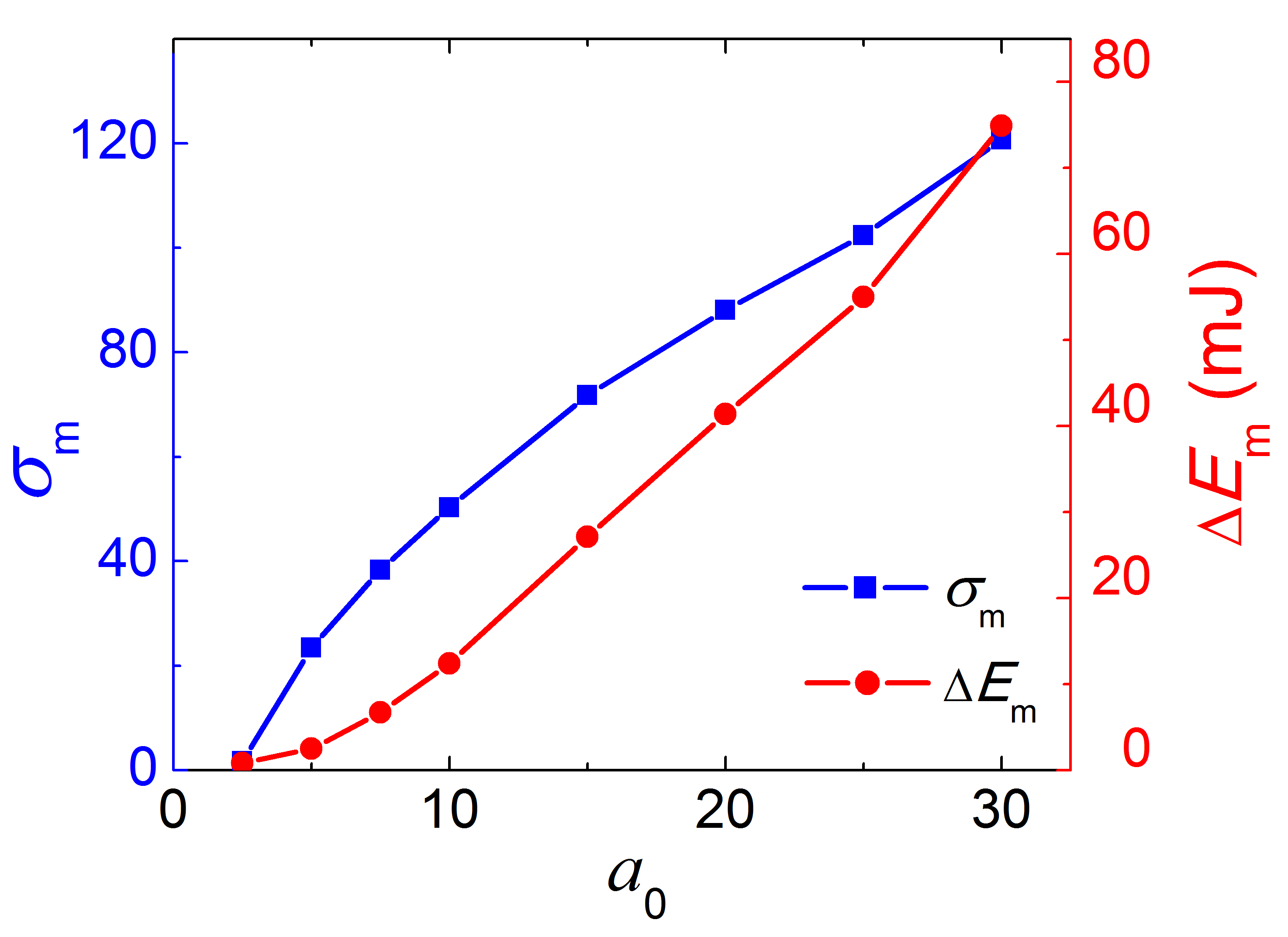}
\caption{\textbf{Relativistic reconnection with various laser
  intensities.} A parameter scan showing the dependence of the maximum magnetized
  parameter $\sigma_{m}$, and the magnetic energy dissipation $\Delta E_{m}$, on
  the normalized laser amplitude $a_{0}$.}
\label{fig:5}
\end{figure}

The present work is the first demonstration that the interaction of a
high-intensity laser and a micro-scale plasma can trigger relativistic
MR, which leads to highly efficient magnetic energy dissipation and
gives rise to intense relativistic jets. In order to explore the
potential of the proposed scenario, we conducted a series of 3D PIC
simulations to study the laser intensity dependence of the maximum
magnetization parameter $\sigma_{m}$ and the magnetic energy
dissipation $\Delta E_{m}$. The results, displayed in Fig.~5 show
that, by applying a micro-sized slab target, relativistic MR can be accomplished
at approximately $a_{0}=2.5$ (where $\sigma_{m} = 1.7$), with a laser
energy of only 50 mJ. Such laser systems are readily available
worldwide, which may open a way to extensive experimental study of
relativistic MR and greatly advance our knowledge of the fundamental
processes such as particle acceleration and magnetic energy
dissipation. Moreover, as the whole setup is micro-scale, the PIC
simulations are considerably less expensive than in previous
laser-plasma driven MR experiments \cite{s46}. Since some processes of interest,
such as the particle acceleration, can only be captured by means of
fully-kinetic PIC simulations, it is therefore more accessible to guide
and interpret future MR experiments via numerical simulations with the
proposed scenario. Nevertheless, it is worth noting that despite the
small physical size ($L\sim1-10\mu m$), due to the high energy
density, in dimensionless variables \cite{s54}
[$L/(\sqrt{\beta}d_{i})\sim10$, where $d_{i}$ is the ion skin depth]
the system here is comparable to previous laser-plasma MR experiments.

Figure~5 illustrates that as the incident laser intensity grows, the
rate at which $\sigma_{m}$ increases gradually slows after the initial
fast-growing phase ($3<a_{0}<10$), while an opposite evolution is
observed for $\Delta E_{m}$. This is because the reconnection site
moves towards the high-density region as the laser-driven electron
beams become increasingly intense, which results in stronger field
dissipation (more electrons gain energy from the reconnecting
field). On the other hand, the enhancement of the local plasma frequency
$\omega_{p}$ slows down the growth of $\sigma_{m}$. Nevertheless, it
is shown that for the normalized laser amplitude $a_{0}=30$
($I\sim10^{21}W/cm^{2}$), the regime of $\sigma\sim 120$ can be
accessed and the released magnetic energy due to field dissipation is
as high as 75 mJ. With next-generation Petawatt laser facilities, such
as ELI \cite{s41}, these results might open a new realm of
possibilities for novel experimental studies of laboratory
astrophysics and highly-efficient particle acceleration induced by
relativistic reconnection.

\section*{Methods}
The 3D PIC simulations presented in this work were conducted with the
code EPOCH \cite{s67}. In this study we restrict the simulations to the
collisionless case, which is justified by the high temperature
($\sim 10 keV$) achieved in laser-plasma interactions, leading to
particles having mean free paths larger than the system size.

A moderately high-intensity laser beam with normalized amplitude
$a_{0} = 5$, and a focus spot of $4\lambda_{0}$ is used to drive the relativistic
MR. The temporal profile of the laser pulse is $T(t) = \sin^{2}(\pi t/\tau)$,
where $0\leq t \leq \tau = 15 T_{0}$, The power and energy of laser pulse
are approximately 12 TW and 200 mJ, respectively. The plasma slab dimension is
$x\times y\times z = 20(7)\lambda_{0}\times1\lambda_{0}\times10\lambda_{0}$,
where the number in the bracket in $x$ direction denotes the length of the
coronal region. The slab is pre-ionized (proton-electron plasma), the initial
temperature is $T_{e}=T_{p}=1$ keV. The plasma density is uniform
($n=n_{0}$) for the main part of the slab, and decreases exponentially
with $x$ in the coronal region, i.e.~$n=n_{0}\exp(-(x-x_{0})^{2}/2l^{2})$,
where $x\geq x_{0}=21\lambda_{0}$ and $l=2\lambda_{0}$ is the scale length.
In most of the simulations we presented in this work [except for Fig.~5],
$n_{0} = 20 n_{c}$ is applied, while in Fig.~5, we use $n_{0} = 50
n_{c}$ for the scan runs in order to avoid problems caused by
self-induced transparency at ultrahigh laser intensities. The
ultra-short laser pulse duration and the relatively-low plasma density are
used to improve computational efficiency, which may cause slight
differences in the simulation outputs, but do not crucially alter the
underlying physics of relativistic MR. For the primary simulation we
presented in this work, the dimensions of the simulation box are
$x\times y\times z = 30\lambda_{0} \times 15\lambda_{0} \times
15\lambda_{0}$, which is sampled by $1200 \times 600 \times 600$ cells
with 5 macro particles for electron species and 3 for proton species
in each cell. The output of this high-resolution simulation was used
to produce Fig.~1--3, and Fig.~4(a). The rest of the results
presented in this work are conducted with a larger simulation box
$x\times y\times z = 40\lambda_{0} \times 20\lambda_{0} \times
20\lambda_{0}$ to ensure the laser pulse and electron beams do not
leave the simulation box while we analyze the magnetic field energy,
but sampled with a lower resolution $800 \times 400 \times 400$ to
reduce the computational time. The numerical convergence has been
confirmed by comparing the physical quantities of interest for the
simulations with different resolutions.


\section*{Supplemental Material}

Here we present a simulation with the setup depicted in Panel~(a) as a
comparison to the results shown in our main paper. In this setup, an
additional plate is placed on the $-y$ side of the plasma slab so that
half of the laser beam is blocked. The plate thickness (in the
longitudinal direction) is $5\lambda_{0}$ and has the same density
($20n_{c}$) as the slab. To ensure that the energy delivered to the
corona [the volume within the blue box frame in Panel~(a)] by the
laser is roughly the same, we increase the laser amplitude by
$\sqrt{2}$ ($a_{0} = 7.07$). In this situation, only one electron beam
reaches the coronal region as shown by the inset in Panel~(a), and as a
result  the effects induced by MR are much weaker.

It should be noted that although almost all of the laser energy on the
$-y$ side of the slab is blocked by the newly-added plate, the surface
current on this side is, as shown by Panel~(b), not exactly
zero. Therefore  MR may still occur in the area with an anti-parallel
magnetic field configuration [zoom-in region in Panel~(b)]. However,
it is found that there are significant differences that support
the arguments that we made in the main paper, and are shown in
Panel~(c-d).

In Panel~(c) we show the field dissipation ($E_{x}j_{x}$) in the corona
in comparison with Fig.~4(a). Obviously, the process is much stronger
and significant in the 2-beam case. The reason is that the topology of
the magnetic field lines are changed  significantly when 2 electron
beams from both sides of the slab reach the corona simultaneously. In
contrast, when there is only one beam (even if it is approximately
twice as intense), the collective motion of electrons, which extract
energy from the collapse of magnetic fields, is almost negligible.

As a result, during the MR process, more magnetic energy is released
and transferred to the plasma in the original 2-beam case, which is
consistent with the analysis of electron energy in the corona, as
shown in Panel~(d). The total kinetic energy gain by the coronal
electrons is 25$\%$ higher in the relativistic MR scenario proposed in
the main paper. Moreover, we note a sharp increase of the
$E_{k,total}$ between 32 $T_{0}$ and 36 $T_{0}$ in the 2-beam case,
which is in good agreement with the period during which the ultrafast
relativistic MR takes place.

\begin{figure}[!b]
\centering
\includegraphics[width=12cm]{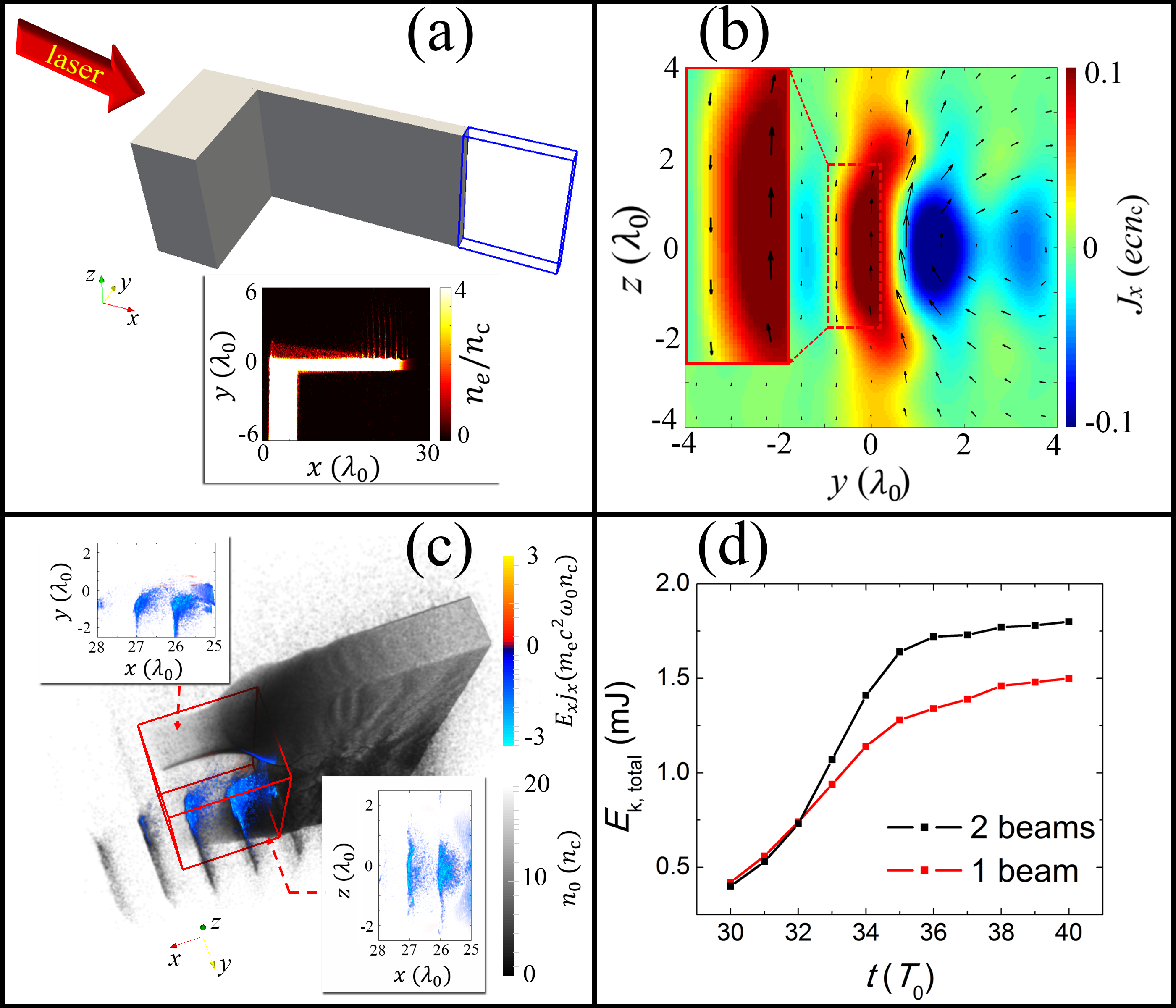}
\caption{\textbf{Schematic and main results from the
  comparison simulation with one laser-driven electron beam.}
  (a) Sketch of the comparison simulation setup, a plate is used to
  block half of the laser pulse so that only one laser-driven
  electron beam reaches the corona (inside the blue-box).
  The inset shows the electron density at cross-section $z = 0$
  and simulation time $t = 30T_{0}$. (b) Longitudinal electric
  current density and transverse magnetic field (black arrows)
  at cross-section $x = 25\lambda_{0}$ and $t = 30T_{0}$, the
  zoom-in area shows the region with anti-parallel magnetic field
  configuration where MR may occur. (c) Field dissipation
  ($E_{x}j_{x}$) and electron density at $t = 33T_{0}$ in the corona
  for the one-beam simulation. (d) Time dependence of total electron
  kinetic energy for one-beam (red) and two-beams (black) cases.
}
\label{fig:s1}
\end{figure}

\section*{Acknowledgements}
The authors would like to thank J Tenbarge, I Pusztai, S Newton, T
Dubois, and the rest of the {\sc pliona} team for fruitful
discussions. This work is supported by the Knut and Alice Wallenberg
Foundation, the European Research Council (ERC-2014-CoG grant 64712)
and National Natural Science Foundation of China (No.11505262). The
simulations were performed on resources at Chalmers Centre for
Computational Science and Engineering (C3SE) provided by the Swedish
National Infrastructure for Computing (SNIC).

\section*{Author contributions statement}
 L.Q.Y. designed and conducted the simulations and analysed the
 results, under supervision of T.F.. L.Q.Y. and T.F. wrote the paper
 with contributions from all the coauthors.

\section*{Additional information}
Competing financial interests: The authors declare no competing
financial interests.

\end{document}